# Real-space topology-engineering of skyrmionic spin textures in a van der Waals ferromagnet Fe$_3$GaTe$_2$

Shuo Mi[1,2,+], Jianfeng Guo[1,2,3,+], Guojing Hu[3], Guangcheng Wang[4], Songyang Li[1,2], Zizhao Gong[3], Shuaizhao Jin[5], Rui Xu[1,2], Fei Pang[1,2], Wei Ji[1,2], Weiqiang Yu[1,2], Xiaolei Wang[4,*], Xueyun Wang[5,*], Haitao Yang[3,*], and Zhihai Cheng[1,2,*]

[1]*Key Laboratory of Quantum State Construction and Manipulation (Ministry of Education) and Department of Physics, Renmin University of China, Beijing 100872, China*

[2]*Beijing Key Laboratory of Optoelectronic Functional Materials & Micro-nano Devices, Renmin University of China, Beijing 100872, China*

[3]*Institute of Physics, Chinese Academy of Sciences, Beijing 100190, China*

[4]*School of Physics and Optoelectronic Engineering, Beijing University of Technology, Beijing 100124, China*

[5]*School of Aerospace Engineering, Beijing Institute of Technology, Beijing 100081, China*

**Abstract:** Realizing magnetic skyrmions in two-dimensional (2D) van der Waals (vdW) ferromagnets offers unparalleled prospects for future spintronic applications. The room-temperature ferromagnet Fe$_3$GaTe$_2$ provides an ideal platform for tailoring these magnetic solitons. Here, skyrmions of distinct topological charges are artificially introduced and spatially engineered using magnetic force microscopy (MFM). The skyrmion lattice is realized by specific field-cooling process, and can be further controllably erased and painted via delicate manipulation of tip stray field. The skyrmion lattice with opposite topological charges (S = ±1) can be tailored at the target regions to form topological skyrmion junctions (TSJs) with specific configurations. The delicate interplay of TSJs and spin-polarized device current were finally investigated via the *in-situ* transport measurements, alongside the topological stability of TSJs. Our results demonstrate that Fe$_3$GaTe$_2$ not only serves as a potential building block for room-temperature skyrmion-based spintronic devices, but also presents promising prospects for Fe$_3$GaTe$_2$-based heterostructures with the engineered topological spin textures.

[+]These authors contributed equally to this work.

[*]Correspondence to Email: zhihaicheng@ruc.edu.cn; htyang@iphy.ac.cn; xueyun@bit.edu.cn; xiaoleiwang@bjut.edu.cn

**Introduction**

Topological magnetic textures have emerged as promising candidates for future spintronic devices, offering unique properties that could revolutionize information storage and processing [1-4]. Among these textures, magnetic skyrmions are nanoscale swirling spin textures exhibiting nontrivial real-space topology [5, 6]. Currently, magnetic skyrmions are being considered for advanced spintronic devices, including racetrack memories, logic gates, and neuromorphic computing, and so on, due to their stability, small size, and controllable responses to external manipulations [7-9]. Furthermore, their low energy-cost current-driven motion favors their potential for practical applications [10-15]. These compelling features present new opportunities for exploring non-trivial topological physics and hold significant promise for future spintronics. To realize skyrmion-based spintronic devices with high integration levels and superior performance, it is crucial to achieve effective control over skyrmion properties, including their size, density, and stability [6,16-18].

Magnetic skyrmions are characterized by parameters such as core polarity, vorticity, and in-plane circular magnetization. Changes of these parameters result in significant variations in skyrmion properties. For instance, skyrmions with different topological charges exhibit contrasting topological Hall effect and skyrmion Hall effect [19-23]. Currently, regulation of skyrmion parameters primarily focuses on macroscopic scales, with overall control achieved through temperature, external magnetic field, and electrical current [24-27]. However, regulation at the microscopic level is lacking. If modulation of the topological order can be achieved locally, unprecedented heterostructures of skyrmions could be realized. This unique magnetic structure may exhibit unexpected electrical properties, thereby advancing development of spintronics.

Two-dimensional (2D) van der Waals (vdW) magnets have recently become attractive platforms for topology-based spintronics due to their fascinating physical properties, such as giant tunneling magnetoresistance and strong spin-orbit coupling. In recent years, materials such as $Cr_2Ge_2Te_6$ [28, 29], $CrI_3$ [30, 31], $Fe_nGeTe_2$ (n = 3, 4, or 5) [32-36] and $CrTe_2$ [37, 38], have been demonstrated to generate magnetic skyrmions. However, most of these materials exhibit a low Curie temperature ($T_C$) below room temperature, limiting their further research and development in practical spintronic devices. Recently, a record-breaking $T_C$ of approximately 350-380 K was discovered in single-crystal Fe$_3$GaTe$_2$, surpassing its extensively studied sister material, Fe$_3$GaTe$_2$. Additionally, recent studies demonstrated that Fe$_3$GaTe$_2$ exhibits ferromagnetism and topological Hall effects at room temperature [39-48], suggesting its potential to host room-temperature skyrmions. The presence of Fe vacancies induces a deviation in the central positions of Fe atoms in Fe$_3$GaTe$_2$, thereby breaking the inversion symmetry and introducing Dzyaloshinskii-Moriya (DM) interactions, ultimately resulting in formation of Néel-type skyrmions [43,49]. Therefore, the discovery of the vdW ferromagnet Fe$_3$GaTe$_2$ offers a promising platform for manipulation of skyrmions at room temperature and provides us with experimental conditions to realize topology engineering of skyrmionic spin textures.

In this work, we demonstrate success manipulation of skyrmions in Fe$_3$GaTe$_2$ using magnetic force microscopy (MFM) under external magnetic fields, achieving unprecedented spatial resolution and control. Firstly, by field cooling (FC), a skyrmion lattice is realized and restricted in a specific range of magnetic fields. Subsequently, controllable painting and erasing of skyrmions is succeeded by delicate manipulation of tip stray field. In particular, by combining magnetic fields, we developed a method for painting skyrmions with different topological charges in Fe$_3$GaTe$_2$, and achieved coexisting skyrmions with opposite topological charges (S = ±1). This painting technique introduces a promising binary bit representation for skyrmion storage devices. Application of the painting technique enabled us to fabricate a unique topological skyrmion junction (TSJ). This TSJ not only enhances electron transmission rates and reduces device resistance, but also serves as an excellent candidate for stacking substrates, capable of providing periodic magnetic fields and topological charges. Finally, the evolution of TSJ properties in magnetic fields reveals that skyrmions with different topological charges exhibit diverse topological properties. Our results not only demonstrate potential applications of Fe$_3$GaTe$_2$ as a storage device, but also contribute to the understanding of topological spin textures.

## Results and discussion

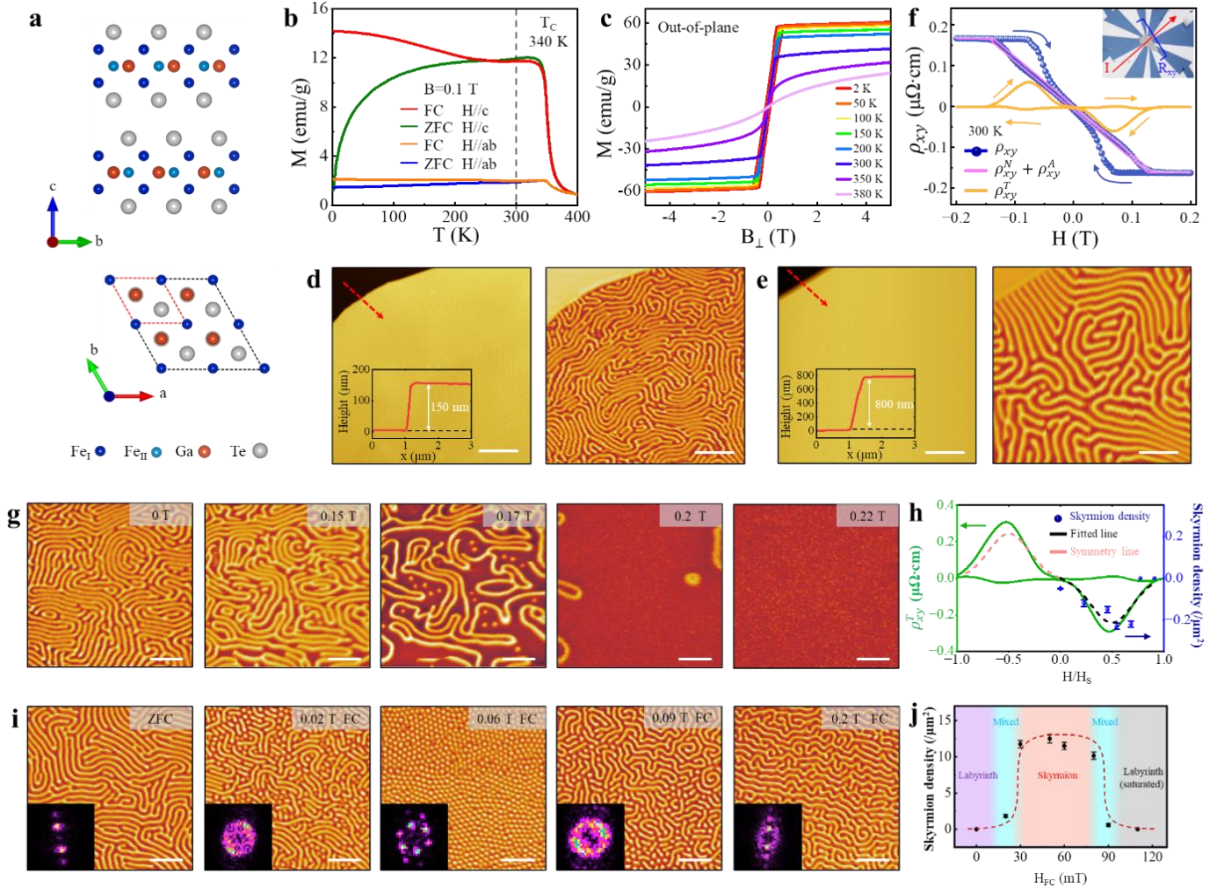

**Figure 1. Structural and magnetic characterization of Fe$_3$GaTe$_2$.** (a) Side- and top-view of the crystal structure. (b) Temperature dependence of zero-field-cooled (ZFC) and field-cooled (FC) magnetizations. (c) Field-dependent magnetizations (*M-H*) at different temperatures with H//c. (d,e) Topography images and corresponding MFM images of Fe$_3$GaTe$_2$ flakes with thicknesses of 150 nm (d) and 800 nm (e). (f) Hall resistivity $\rho_{xy}$, normal and anomalous Hall resistivity $\rho_{xy}^N + \rho_{xy}^A$, and topological Hall resistivity $\rho_{xy}^T$ as functions of *H* for a Fe$_3$GaTe$_2$ flake (200 nm) at 300 K. (g) MFM images of Fe$_3$GaTe$_2$ flake with increasing magnetic fields at room temperature. The magnetic field is parallel to *c* and considered to be positive in the upward direction. (h) $\rho_{xy}^T$ and skyrmion density derived from (g) versus *H* for the flake. The black dashed line represents the fit curve derived from the skyrmion density, while the pink dashed line is derived from the center inversion. (i) MFM images obtained at 0 T after different FC. The insets show the corresponding FFT maps. (j) Skyrmion density curve obtained from data fits in (i), showing different magnetic phases. Scale bar: 2 μm.

Fe₃GaTe₂ is a member of the vdW ferromagnet family, in which each layer consists of a Fe₃Ga layer sandwiched by Te layers, as shown in Fig. 1a. The crystal structure of Fe₃GaTe₂ corresponds to the P6$_3$/mmc space group, which exhibits centrosymmetry with lattice parameters a = b ≈ 3.99 Å and c ≈ 16.23 Å [40]. The M-T curves reveals that the T$_C$ of Fe₃GaTe₂ reaches 340 K, surpassing that of all known vdW ferromagnets, as shown in Fig. 1b. The variable temperature M-H curves confirm the high T$_C$ of Fe₃GaTe₂ and demonstrate that an easy magnetization axis aligns with the crystalline c-axis, as shown in Fig. 1c and Fig. S1. In addition, the X-ray diffraction pattern and Raman spectra demonstrate high quality of the single crystals used in our experiments (Fig. S1). The Fe₃GaTe₂ flakes obtained through mechanical exfoliation were characterized by MFM at room temperature, as shown in Fig. 1d,e. The real space iamges show that Fe₃GaTe₂ exhibits a FM labyrinthine domains in the ground state, and the domain width is clearly influenced by thickness. Specifically, the domains width increases with the thickness of Fe₃GaTe₂ flakes, following Kittel's law (Fig. S2).

The EDS spectrum reveals an approximate Fe:Ga:Te ratio of 2.8:1:1.9, indicating the presence of Fe deficiency in the flakes (Fig. S1). According to previous reports [49], the presence of Fe deficiency in Fe₃GaTe₂ introduces DM interactions, which results in the formation of topological spin textures. To verify this, we performed Hall measurements on the flake, as shown in Fig. 1f. A significant presence of topological Hall resistance ($\rho_{xy}^T$) is observed, therefore confirming the existence of topological spin textures. Further MFM measurements confirm that the topological spin textures in Fe₃GaTe₂ correspond to magnetic skyrmions, as shown in Fig. 1g. The increase in magnetic field leads to a decrease of labyrinthine domains in Fe₃GaTe₂, accompanied by the generation of skyrmions. The fitting curve of skyrmion density, obtained from the MFM image, exhibits excellent agreement with the $\rho_{xy}^T$ curve, as shown in Fig. 1h.

To achieve a uniform skyrmion lattice, we conducted field-cooling experiments at various magnetic fields on the Fe₃GaTe₂ flake and subsequently performed MFM measurements, as shown in Fig. 1i. The characterization revealed the presence of three distinct magnetic phases, namely labyrinthine domain, mixed phase, and skyrmion lattice.

Among them, the skyrmion lattice manifest within the range of 0.03 T-0.08 T (Fig. S3). The phase diagram of $Fe_3GaTe_2$ was constructed based on the experimental results, as shown in Fig. 1j. In addition, we also examined the skyrmion lattices in $Fe_3GaTe_2$ flakes with varying thicknesses, and it is evident that the size of the skyrmion depends on the thickness (Fig. S4). The exceptional storage density and high stability properties of magnetic skyrmions position them at the forefront of promising candidates for building next-generation information storage devices [50-52]. For a truly functional storage device, skyrmions must be able to created and erased in a deterministic manner. In order to demonstrate the potential of $Fe_3GaTe_2$ as a next-generation storage device, we conducted further investigations into manipulating the topological spin texture in $Fe_3GaTe_2$ and successfully achieved the implementation of skyrmion painting and erasing.

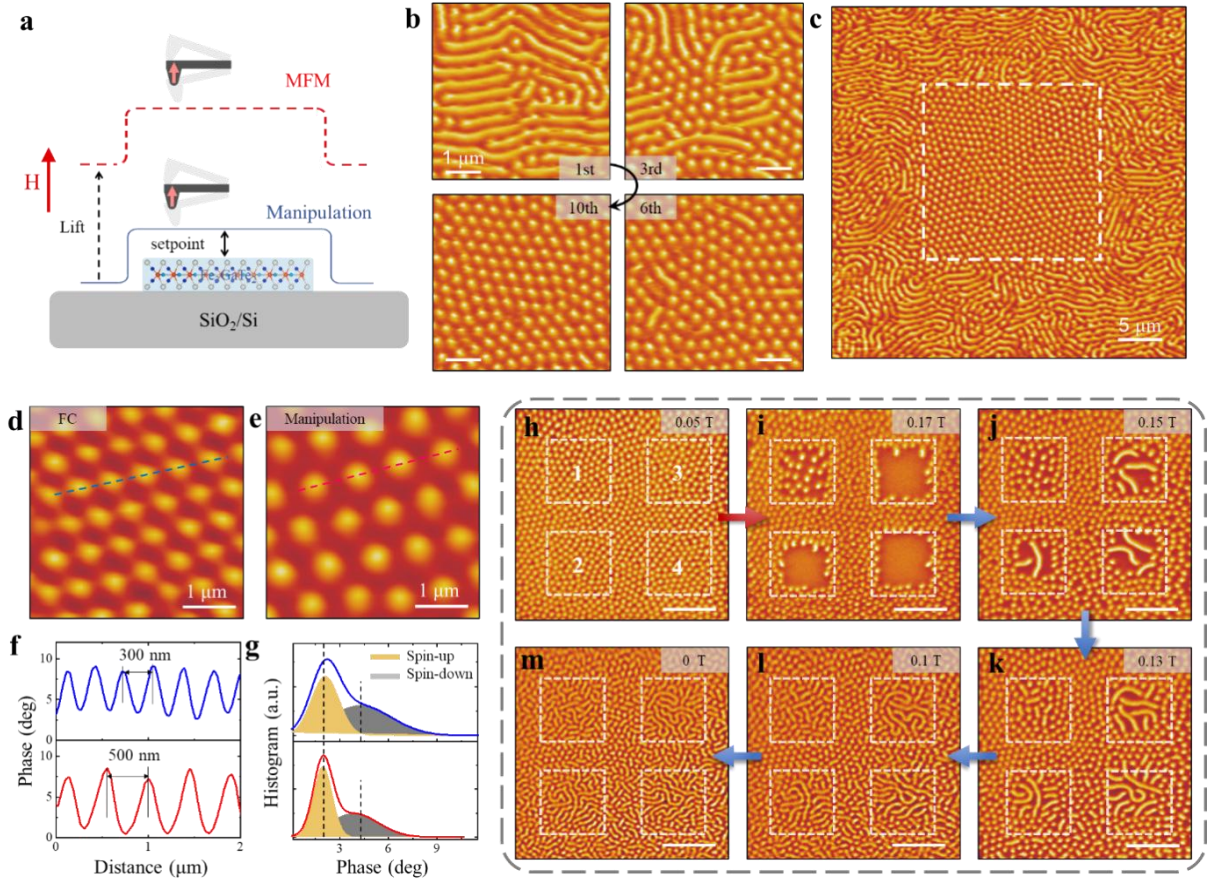

**Figure 2. Formation and erasure manipulations of skyrmions on Fe$_3$GaTe$_2$ flake (200 nm) at 300 K.** (a) Schematic diagram of manipulating skyrmions using a magnetic tip. (b) MFM images of the skyrmions generation processed at 0.05 T, corresponding to 1st, 3rd, 6th and 10th manipulations, respectively. (c) Large-scale MFM image after manipulating the generation of skyrmions. The manipulation area is indicated by the white dashed box. (d,e) Skyrmions generation by FC (d) and by manipulation (e) under a constant field of 0.05 T. (f) Line profiles along the blue (d) and red (e) dashed lines. (g) Histograms of MFM signals for (d) and (e). (h,i) MFM images before (h) and after (i) erasing skyrmions. The white dashed line delineates the manipulation region and marks the number of manipulations. (j-m) MFM images with decreasing magnetic fields after erasing skyrmions. Scale bar: (b,d,e) 1 μm; (c,h-k) 5 μm.

The MFM setting illustrated in Fig. 2a demonstrates the manipulation of the magnetic structures within a $Fe_3GaTe_2$ flake, where skyrmions are painted and erased by interacting with the stray field produced by the magnetic tip. The stray field of the magnetic tip may reach magnitudes in the thousands of Gauss and exhibits rapid decay along the z-direction (Fig. S5d) [16]. Therefore, by adjusting the setpoint, we can effectively utilize MFM as either characterization or manipulation modes (Fig. S5). In the manipulation mode, a uniform skyrmion lattice is successfully painted through the combination of the stray field and an applied magnetic field (0.05 T), as shown in Fig. 2b. The density of the skyrmion can be varied with the number of manipulations. Performing a single manipulation results in a lower quantity of skyrmions, while seven manipulations essentially forms a skyrmion lattice with a nearly hexagonal structure (Fig. S6). It is worth noting that written skyrmions remains stable below the saturation field and is unaffected by the external static magnetic field. The large-scale MFM image in Fig. 2c displays a stable and homogeneous skyrmion lattice in the manipulation region, while the surrounding magnetic domains remain primarily in a labyrinthine domain. The clear boundary indicates the high precision of our manipulation.

Skyrmion lattice generated through FC and painted through manipulation are further compared, as shown in Fig. 2d-g. The distance between the manipulation-written skyrmions is slightly longer and exhibits higher homogeneity, forming an almost perfect hexagonal lattice when compared with the skyrmions created by simple FC. Furthermore, the MFM signals of each skyrmion generated in both methods exhibit minimal differences (Fig. 2g), indicating their identical topological nature (vorticity). Afterwards, we proceeded to manipulate the skyrmion lattice by erasing the skyrmions successfully at 0.17 T, as shown in Fig. 2h,i. The numerical value within the central white dashed box indicates the quantity of manipulations. The skyrmions can be partially erased with a single manipulation, while complete disappearance of skyrmions occurs when the number of manipulations exceeds two. The subsequent MFM characterization with decreasing magnetic field reveals that in regions with a low density of skyrmions, the descendance of magnetic field induces the formation of labyrinthine domains and destabilizes the surrounding skyrmions, as shown in Fig. 2i-m.

So far, the painting and erasing of the skyrmion in $Fe_3GaTe_2$ have been successfully achieved without relying on thermodynamics, thereby confirming the potential of $Fe_3GaTe_2$ as a next-generation storage device. With the hexagonal lattice structure, the presence or absence of skyrmions can be utilized as a binary bit of 1 or 0, respectively. Note that absence of skyrmions in certain regions may lead to formation of labyrinthine domains, which can negatively impact the performance of devices fabricated on $Fe_3GaTe_2$.

The highly manipulatable nature of the skyrmion lattice in $Fe_3GaTe_2$ prompts us to propose a novel approach for binary bit representation, i.e., the utilization of skyrmions with different topological charges. The following section presents our topology-engineering of skyrmions in $Fe_3GaTe_2$. We have successfully achieved the unprecedented coexistence of skyrmions with distinct topological charges and constructed a specific magnetic texture with unique electrical properties.

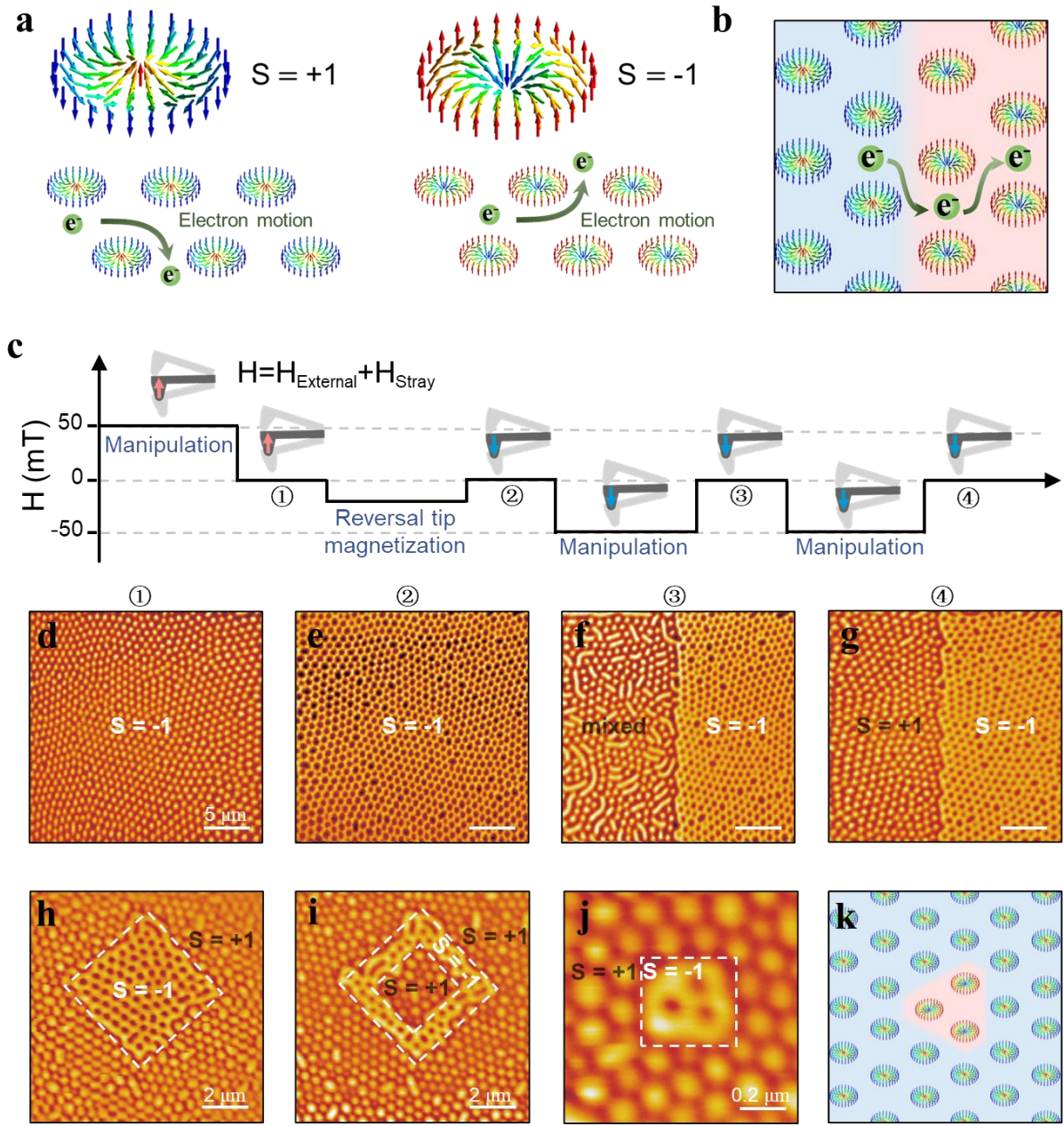

**Figure 3. Topology-engineering of skyrmions on Fe$_3$GaTe$_2$ flake.** (a) Schematic illustration of skyrmions with different topological charges (S = ±1). The lower insets show the motion of electrons influenced by skyrmions. (b) Schematic drawing of electron motion at the boundary of the skyrmion lattice with different topological charges. (c) Schematic drawing of the manipulation processed for realizing coexisting skyrmions with different topological charges. (d-g) Corresponding MFM images in (c). (h-j) MFM images after topology-engineering of skyrmions. The fine patterning of distinct topological skyrmions can be accomplished by using a magnetic tip. (k) Schematic drawing of different topological skyrmions in (j). Scale bar: (d-g) 5 μm; (h,i) 2 μm; (j) 0.2 μm.

Skyrmions can, indeed, be defined by the topological charge S (or skyrmion number), which is a measure of the winding of the normalized local magnetization (m). In the two-dimensional limit, the topological number is [6, 26, 53]:

$$S = \frac{1}{4\pi} \int \bm{m} \cdot (\partial_x \bm{m} \times \partial_y \bm{m}) dx dy = \pm 1$$

The normalized magnetization can be mapped on a unit sphere and, in the case of skyrmions, it covers the entirety of the sphere ($4\pi$) and is thus quantized. In $Fe_3GaTe_2$, skyrmions exhibit a Néel-type configuration, characterized by radial-shaped spin textures and possessing zero helicity [43]. These topologically non-trivial chiral spin textures are characterized by topological charges $S = \pm 1$. The spin configurations with topological charges $S = \pm 1$ is illustrated in Fig. 3a. Due to their non-trivial nature, skyrmions induce a topological Hall effect by acting as emergent electromagnetic fields, deflecting the motion of electrons. It is worth noting that the motion directions of electrons in $S = 1$ and $S = -1$ skyrmion lattices are opposite. Therefore, if electrons can traverse regions with opposing topological charges of skyrmions, their deflection of motion is effectively counteracted [4, 26, 54], thereby significantly enhancing the rate of longitudinal electron conductivity, as illustrated in Fig. 3b.

The controllable generation of skyrmions with different topological charges can be achieved by reversing the magnetic moment of the magnetic tip, as shown in Fig. 3c. The first step involvs creation of a skyrmion lattice with $S = -1$ in $Fe_3GaTe_2$, as illustrated in Fig. 3d. Subsequently, a weak magnetic field of approximately 0.01 T is applied to reverse the moment of the magnetic tip, resulting in a corresponding signal reversal in the MFM image, as shown in Fig. 3e. Following this, manipulation was performed in a specific region (left rectangle) of the skyrmion lattice with $S = -1$. After a single manipulation, it is evident that $S = 1$ skyrmions are generated in the manipulation region, as shown in Fig. 3f. Following multiple manipulations, the manipulated region undergoes a complete transformation into an $S = 1$ skyrmion lattice, resulting in coexisting skyrmion lattices with distinct topological charges, as shown in Fig. 3g. This coexisting phase is stable and exhibits stability at room temperature even in the absence of an external magnetic field. We then named this special magnetic structure as a "topological skyrmion junction (TSJ)".

More detailed exploration of this method for painting skyrmion lattices with different topological charges reveals its high level of manipulability and precision. The manipulability of this painting method enables realization of any desired complex pattern on $Fe_3GaTe_2$. Through a systematic experimental design, we successfully achieved intricate patterning on the skyrmion lattice of $Fe_3GaTe_2$, as shown in Fig. 3h,i. The high precision of this painting method enables us to manipulate the magnetic texture more finely, even down to several individual skyrmions. Fig. 3j,k shows the MFM image and corresponding schematic drawings after writing three skyrmions with opposite topological charges. These results validate extensive applicability of this painting method, thereby offering novel insights and platform for future development and investigation in storage devices based on skyrmions.

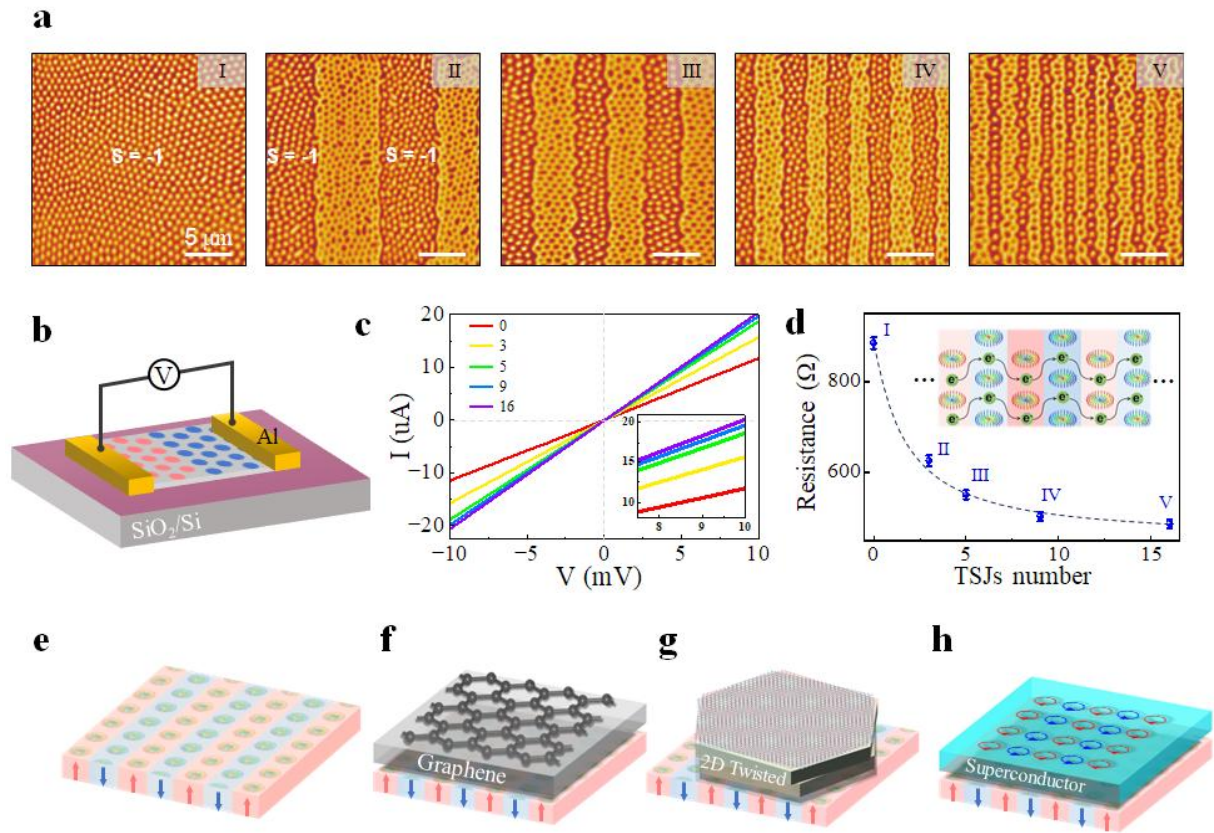

**Figure 4. TSJ and resistance modulation action.** (a) MFM images with different number of TSJs. (b) Schematic illustration of the resistance measuring device. (c,d) Current-voltage (I-V) curves (c) and resistance (d) of periodic skyrmion arrays in (a). The numerical values correspond to the number of TSJs present in different periodic skyrmion arrays. The inset in (d) shows the motion of electrons in TSJs. (e) Schematic diagram of TSJs with rows of single skyrmion. This particular topological magnetic structure exhibits spatially periodic magnetic fields and topological charges. (f-h) Schematic of stacking graphene (f), Moiré pattern (g), and superconductors (h) on skyrmion arrays shown in (e). Scale bar: 5 μm.

Theoretically, as discussed previously, TSJ can significantly enhance the electron transmission efficiency, thereby effectively reducing the resistance of the sample. The proposed novel method for manipulating skyrmions offers an opportunity to artificially construct TSJs. Following this, a series of periodic skyrmion arrays were constructed and their MFM images are depicted in Fig. 4a. Notably, the minimum limit, that is, a single skyrmion periodic array, has been achieved. The periodic skyrmion arrays are distinguished by the number of TSJs present, which corresponds to 0, 3, 5, 9, and 16 from left to right. To investigate the electrical properties of these periodic skyrmion arrays, resistance measurements were conducted on them using the experimental setup illustrated in Fig. 4b. The measured current-voltage (I-V) curves in Fig. 4c demonstrate a noticeable increase in slope with the number of TSJs. By extracting the resistance of the various periodic skyrmion arrays based on these I-V curves, the relationship between resistance and the number of TSJs is depicted in Fig. 4d. The resistance decreases as the number of TSJs increases, in accordance with expectations. The above experimental results demonstrate that the implementation of TSJ can effectively enhance electron transmission rates, thereby resulting in reduced energy consumption and improved efficiency of electronic devices.

In addition to their unique electrical properties, TSJs are also an excellent candidate for bottom substrates that modulates properties of other samples through proximity effects. This particular topological magnetic structure not only facilitates presence of periodic topological charges, but also enables generation of periodic magnetic fields, as shown in Fig. 4e. Considering the challenge and importance of realizing spatial controllable periodic magnetic fields [55], our findings provide an opportunity to explore novel magnetic phenomena and their potential applications with the bottom substrate. For instance, by modulating the electronic states and band structures in two-dimensional materials such as graphene (Fig. 4f) or moiré superlattices (Fig. 4g), it is possible to achieve intricate topological quantum states, and pave the route for development of quantum devices with exceptional properties [56-59]. Moreover, manipulating superconducting pairing channels in superconductors (Fig. 4h) allows for production of unconventional or topologically non-trivial superconducting phases [60, 61].

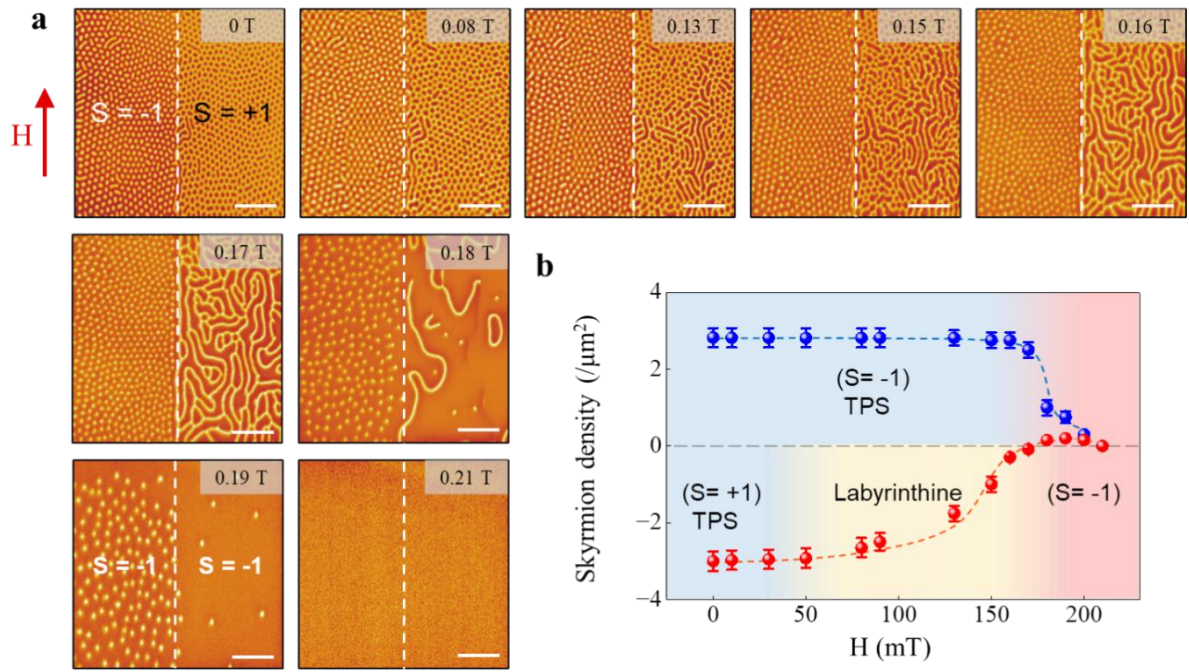

**Figure 5. Topological properties of skyrmions with different topological charges.** (a) MFM images of the TSJ with increasing magnetic fields. (b) The variation curve of skyrmion density with magnetic field in S = +1 and S = -1 regions. Scale bar: 4 μm.

Finally, we investigate the temperature stability and topological properties of TSJ in $Fe_3GaTe_2$. The MFM images obtained during the cooling process reveal that the TSJs exhibit variable temperature stability and can be effectively stabilized down to at least 200 K (Fig. S7). Further magnetic field-dependent MFM characterization reveals differences in the topological properties of skyrmions with different topological charges, as shown in Fig. 5a. For skyrmion with S = -1, which is antiparallel to the magnetic field, remains stable below the saturation field and exists in a topologically protected state (TPS). Skyrmions gradually vanish near the saturation field as they are unable to maintain their spin texture, ultimately transitioning into the saturated state. In contrast, for the skyrmion with S = 1, which is parallel to the magnetic field, its spin texture is more susceptible to disruption caused by magnetic field, and the TPS remains stable only up to 0.05 T. Subsequently, the skyrmions undergo fusion and expansion, resulting in compression of the background region into a labyrinthine domain. The labyrinthine domains undergo further fragmentation and contraction under the influence of the magnetic field, ultimately transforming into S = -1 skyrmions near the saturation field before eventually reaching the saturation state. The difference in the topological properties of the S = 1 and S = -1 skyrmion lattices can be intuitively visualized by examining the skyrmion density as functions of fields, as shown in Fig. 5b.

**Conclusions**

In summary, our study highlights the potential of a 2D vdW ferromagnet $Fe_3GaTe_2$ as a platform for investigating and engineering topological textures with magnetic skyrmions. We found that painting and erasing of skyrmion lattices were successfully achieved by using a magnetic tip, eliminating the necessity of thermal cycling (FC) . We also successfully realized both S = 1 and S= -1 skyrmion lattices, as well as coexistence of both type of skyrmions. As an example, we fabricated a unique TSJ and found that it improves electron transmission rates with reduced device resistance. Finally,  skyrmions with opposite topological charges exhibiting different field-dendence of skyrmiosn densities. Our results provide topological skyrmions as candidate palforms for development of future storage devices with binary representation and  high-efficiency transport devices, and serving as a bottom substrate with spatially periodic magnetic fields, which opens new route for investigating topological spintronics and new quantum states.

**Acknowledgments**

This project is supported by the National Key Research and Development Projects of China (No. 2023YFA1406500 and No. 2022YFA1204100), Strategic Priority Research Program (Chinese Academy of Sciences, CAS) (No. XDB30000000, XDB33030100), National Natural Science Foundation of China (NSFC) (No. 61674045, 61911540074, 62488201, 12074018, 92163101, 12374080), the Innovation Program of Quantum Science and Technology (2021ZD0302700), and Fundamental Research Funds for the Central Universities and Research Funds of Renmin University of China (No. 21XNLG27).

## Materials and Methods

### Crystal and sample preparation

High-purity Fe power (99.99%), Ga lumps (99.99%) and Te power (99.999%) were mixed at atomic ratios of 1:1:2 in a sealed quartz tubes with high vacuum, in which the Ga and Te are also acting as the flux. The quartz tubes are placed in a muffle furnace and heated to 1273 K in 1 h. After holding for 24 h, the temperature is rapidly lowered to 1153 K in 1 h. Then, the temperature is slowly lowered to 1053 K over a period of 120 h. Finally, single crystal of $Fe_3GaTe_2$ can be obtained by centrifugation.

### Device fabrication and Hall measurement.

Thin flakes of $Fe_3GaTe_2$ were obtained via mechanical exfoliation from synthetic bulk crystals onto $SiO_2$(300 nm)/Si substrates transferred using polydimethylsiloxane (PDMS). Hall bar devices were fabricated by photolithography and magnetron sputtering, with $Fe_3GaTe_2$ flake positioned on the four Au electrodes to form a 34 μm wide and 10 μm long channel for perpendicular voltage detection. Four-probe electrical measurements of transverse/Hall resistance ($R_{xy}$) dependent on magnetic field were conducted using an electromagnet system, with a fixed 0.1 mA DC current applied to the channel.

### Resistance measurement.

The resistance measurements were performed by double channel Keithley 2400 SourceMeter with a two-probe station system. All resistance measurements (with varying numbers of TSJs) are captured on the same $Fe_3GaTe_2$ flake.

### AFM and MFM measurement

The AFM and room temperature MFM experiments were conducted at room temperature using a commercial magnetic force microscope (Park NX10) equipped with a commercial magnetic tip (Nanosensors, PPP-MFMR). The scanning probe system operated at the resonance frequency of the magnetic tip (approximately 75 kHz). MFM imaging was performed in standard two-pass tapping mode, acquiring topography during the first pass and MFM signal during the second pass at a set lift height of nominally 100 nm. The manipulations and cryogenic MFM experiments were conducted using a commercial magnetic force microscope (attoAFM I, attocube) employing a commercial magnetic tip (Nanosensors, PPP-MFMR) based on a closed-cycle He cryostat (attoDRY2100, attocube). The manipulation of labyrinthine domains was performed in tapping mode, where reducing the setpoint value of the tip amplified the impact of the stray field, thereby facilitating precise generation of skyrmions.